\pdfoutput=1

\documentclass[aip,jcp,reprint,floatfix]{revtex4-2}

\usepackage{amsmath}
\usepackage{graphicx}
\usepackage{hyperref}
\usepackage{orcidlink}

\usepackage{adjustbox} % allow fitting of overfull items within page width

\begin{document}

\title{Dissociation of Adsorbates via Electronic Energy Transfer from Aromatic Thin Films}
\author{E.T. Jensen \\
Department of Physics\\
University of Northern BC\\
\href{https://orcid.org/0000-0001-8030-4204}{\includegraphics[scale=0.5]{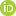}} {orcid.org/0000-0001-8030-4204}}

\date{October 4, 2024}

\begin{abstract}

Photofragment translational spectroscopy has been used to characterize the photodissociation of CH$_3$I and CF$_3$I adsorbed on thin films of a variety of aromatic molecules, initiated by near-UV light. Thin films (nominally 10 monolayers) of benzene, five substituted benzenes and two naphthalenes have been employed to study systematic changes in the photochemical activity. Illumination of these systems with 248nm light is found to result in a dissociation process for the CH$_3$I and CF$_3$I mediated by initial absorption in the aromatic thin film, followed by electronic energy transfer (EET) to the dissociating species. The effective cross sections for dissociation are found to be substantially increased via this mechanism, by amounts that differ depending on the aromatic molecule thin film used, and is connected to the aromatic photabsorption profile. Distinctive translational energy distributions for the CH$_3$ and CF$_3$ photofragments are found to vary systematically for the different aromatic molecule thin film used, and are related to the aromatic molecule excited states. The CH$_3$ and CF$_3$ photofragment kinetic energy distributions found for the aromatic thin films suggest that the dissociation occurs via EET to the $^3Q_1$ excited state of CH$_3$I and CF$_3$I.

\end{abstract}

\maketitle

\section{Introduction}

In surface photochemistry the tools of surface science are employed to create and characterize surface and adsorbate systems to be studied. In so doing, a variety of photochemical mechanisms have been identified as responsible for processes in adsorbed systems, such as photodissociation, photodesorption and photoreactions\cite{zhou:1991,hasselbrink:2008}. The richness of organic photochemistry seen in aromatic molecular systems has been observed in some surface photochemical systems, such as photodesorption observed from UV irradiation of benzene/water ice\cite{thrower:2008} demonstrating that electronic excitation of benzene leads to desorption of benzene and water molecules. Such processes have been of active interest for atmospheric chemistry and astrochemistry, where non-thermal excitations (in this case low energy UV photons) can lead to chemical changes and desorption of species in mixed systems of molecular ices.\cite{fillion:2014, bertin:2023}.

In order to predict the outcomes from the processes that can be involved in the complex systems such as those found in atmospheric chemistry\cite{grannas:2007,finlayson-pitts:2009} and astrochemistry\cite{Hornekr:2014,oberg:2016,mullikin:2018,arumainayagam:2019}, knowledge of the fundamental processes and mechanisms, their cross sections and their interactions with various species needs to be understood. The present work examines electronic energy transfer in a variety of aromatic molecule thin films, in which EET to coadsorbed CH$_3$I or CF$_3$I molecules is used as a probe. The consequent dissociation of the CH$_3$I or CF$_3$I is  observed using photofragment translational spectroscopy to characterize the energetics of the dissociation process and the properties of the aromatic thin film that mediates the transfer of the initial photoexcitation to the dissociating coadsorbate.

To further characterize the EET dissociation mechanism identified for CH$_3$I on thin films of fluorobenzenes and  benzene\cite{jensen:2024,jensen:2021}, in the present work we have extended our study to examine thin films of several mono-substituted benzenes (Section {\ref{monobenz_section}}), two naphthalenes (Section {\ref{naph_section}}), as well as for CF$_3$I adsorbed on several of these aromatic thin films (Section \ref{CF3I_pdissn}).

\subsection{Near-UV Photoabsorption by Aromatics}
Small aromatic molecules have well-characterized near-UV absorption from the ground-state to excited bound states with vibronic structures\cite{turro:2010}. For benzene, the lowest singlet state $S_1$ ($^1\!B_{2u}$ excitation) is due to a $\pi$--$\pi^*$ transition in the wavelength region 230--262nm (4.75--5.40eV). For naphthalene, the $S_1$ state is lower in energy ($^1\!B_{3u}$, 315--295nm; 3.94--4.20eV) due to a $\pi$--$\pi^*$ `long-axis' transition, and the second excited singlet ($S_2$) state `short axis' transition ($^1\!B_{2u}$) in the range 245--290nm (4.28--5.06eV). These vibronic transitions are broadened and slightly red-shifted in the adsorbed state (for benzene, by 1.8nm\cite{dawes:2017}, for naphthalene $\sim$5nm\cite{craig:1961}). The mono-substituted benzenes and naphthalenes studied in the present work display photoabsorption to analogous states in the same energy region as unsubstituted parents, typically at slightly lower energies. 

Near-UV excitations in small aromatic molecules can be followed by various outcomes. The $S_1$ states tend to have sufficiently long lifetimes that internal conversion will populate lower vibronic states in the $S_1$ manifold, fluoresce to return to the ground state ($S_0$), transition via an inter-system crossing (ISC) to lower-energy triplet states ($T_1$ or $T_2$) or undergo non-radiative transitions to the $S_0$ vibronic manifold. If the initial absorption is to a higher singlet state (e.g. $S_2$), most aromatics undergo rapid internal conversion to the longer-lived $S_1$ state, which is referred to as `Kasha's Rule'\cite{turro:2010}. Fluorescence from the $S_1$ vibronic states back to the ground state $S_0$ manifold is at lower photon energies than for photoabsorption, beginning from the $S_1$ 0-0 edge. For example, in gas-phase benzene the $S_1$ fluorescence band extends from 262--315nm (3.94--4.75eV) and for naphthalene from 312--410nm (3.01--3.97eV). Table {\ref{table_1_energies}} summarizes the $S_1$ 0-0 band origin energies for the aromatics studied in the present work.

\begin{table*}[tpb]
\caption{Energies of the $S_1$ 0-0 band origins for aromatic molecules used for thin films in the present work reported for condensed-phase systems, or for the gas-phase (in brackets) if no condensed phase data is available. The observed CH$_3$ photofragment peak kinetic energies from CH$_3$I EET dissociation on these thin films are shown in the second row. The final row shows the estimated EET excitation energy (E$_{\textrm{exc}}$ in Equ. {\ref{Equ_2}}) for each thin film. }
%\begin{adjustbox}{width={\textwidth},totalheight={\textheight},keepaspectratio}

\begin{tabular}{|lcccccccc|}
\hline
\multicolumn{1}{|l||}{}   & \multicolumn{1}{c|}{C$_6$H$_6$}   & \multicolumn{1}{c|}{C$_6$H$_5$F}   & \multicolumn{1}{c|}{1,4-C$_6$H$_4$F$_2$}   & \multicolumn{1}{c|}{C$_6$H$_5$CH$_3$}   & \multicolumn{1}{c|}{C$_6$H$_5$OH}   & \multicolumn{1}{c|}{C$_6$H$_5$CCH}   & \multicolumn{1}{c|}{C$_{10}$H$_8$}   & 1-C$_{10}$H$_7$F   \\
\hline
\multicolumn{1}{|l||}{S$_1$ Energy (eV)}  & \multicolumn{1}{c|}{4.69\cite{dawes:2017}}  & \multicolumn{1}{c|}{(4.69)\cite{fukuzumi:1991}}  & \multicolumn{1}{c|}{(4.57)\cite{fukuzumi:1991}}  & \multicolumn{1}{c|}{(4.65)\cite{serralheiro:2015}}  & \multicolumn{1}{c|}{(4.51)\cite{limao-vieira:2016}}  & \multicolumn{1}{c|}{4.41\cite{swiderek:1998}}  & \multicolumn{1}{c|}{3.91\cite{craig:1961}}  & (3.95)\cite{singh:1981}  \\
\hline
\multicolumn{1}{|l||}{CH$_3$ KE (eV)} & \multicolumn{1}{c|}{1.70} & \multicolumn{1}{c|}{1.55} & \multicolumn{1}{c|}{1.45} & \multicolumn{1}{c|}{1.45} & \multicolumn{1}{c|}{1.45} & \multicolumn{1}{c|}{1.45} & \multicolumn{1}{c|}{1.12} & 1.12 \\
\hline
\multicolumn{1}{|l||}{E$_{\textrm{exc}}$ (eV)} & \multicolumn{1}{c|}{4.52} & \multicolumn{1}{c|}{4.35} & \multicolumn{1}{c|}{4.24} & \multicolumn{1}{c|}{4.24} & \multicolumn{1}{c|}{4.24} & \multicolumn{1}{c|}{4.24} & \multicolumn{1}{c|}{3.88} & 3.88 \\
\hline
\end{tabular}
%\end{adjustbox}

\label{table_1_energies}
\end{table*}

Electronic energy transfer (EET) in molecular systems is a phenomenon found in a variety of contexts, where the long-lived excitation in a molecule is transferred to a nearby molecule. For example, gas-phase photosensitization resulting in enhanced UV photodissociation of CH$_3$I in mixtures with benzene vapour was noted over 70 years ago\cite{dubois:1951}. There are many heteromolecular systems where the role of EET has been studied in detail via fluorescence (e.g. mixed dimers of p-difluorobenzene:p-xylene\cite{lahmani:1990} and benzene:biacetyl\cite{bigman:1994}). In condensed-phase systems of aromatic molecules the transport of localized electron–hole excitations (homomolecular EET) is described in terms of Frenkel excitons (electron+hole on the same molecular site) or charge-transfer (CT) excitons when the electron–hole pair are separated on nearest-neighbour sites.\cite{bardeen:2014} 

The lifetimes of the excitation are a property of relevance for the EET mechanism. In low-temperature (T\textless 200K) benzene crystals, the $S_1$ fluorescence lifetime is found to be $\sim$80ns\cite{Lumb:1971}. For gas-phase fluorobenzene the $S_1$ fluorescence lifetime is  shorter (10-20ns)\cite{phillips:1972} than that of benzene, which is a general trend the substitute benzenes. For many of the substituted benzenes, the fluorescence lifetimes in low-temperature condensed phases is not available in the literature. For crystalline naphthalene, the fluorescence lifetime is $\sim$100ns\cite{auweter:1979}. In the present work the excited state lifetime is of significance if this electronic excitation is to be transported within the aromatic crystal thin film and/or transferred to the CH$_3$I or CF$_3$I molecules adsorbed on top of the aromatic film.

\subsection{Near-UV Photodissociation and Excited States of Methyl Iodide}
\label{CH3X_pdissn}
\begin{figure}[b]
% Fig 1 
% CH3I PES from Aleksyev
\includegraphics[scale=0.75]{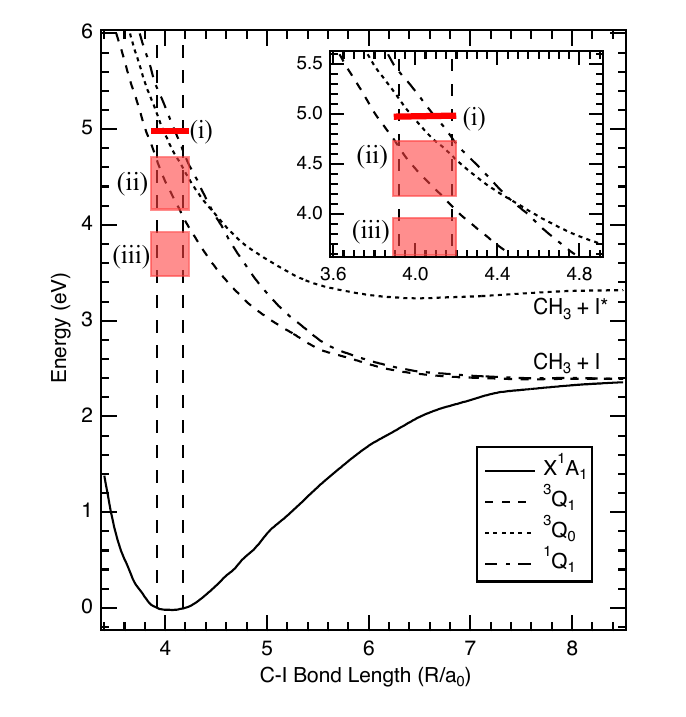}
\caption{Potential energy curves for the ground state, and dissociative states that comprise the A-band of the CH$_3$I monomer, based on data from Ref. {\onlinecite{alekseyev:2007}}. The zero of energy is set at the ground vibrational state of the C--I bond. The vertical dashed lines indicate the Franck-Condon region for ground-state CH$_3$I. The inset shows detail for the energy region where the A-band excitations occur. The curve-crossing between the $^3Q_0$ and $^1Q_1$ states occurs outside the Franck-Condon region. The highlighted regions indicate the energy regions for: (i) 248nm photons; (ii) $S_1$ fluorescence band of mono-substituted benzenes; and (iii) $S_1$ fluorescence band of naphthalenes. }
\label{fig_1_CH3I_PES}
\end{figure}

Photodissociation of gas-phase CH$_3$I in the near-UV region occurs via the `A-band', a set of $n\rightarrow\sigma^*$ transitions observed as three overlapping states between $\sim$3.5eV and 5.6eV ($^3Q_1$, $^3Q_0$ and $^1Q_1$ in order of increasing energy) above the ground state in the Franck-Condon region\cite{Eppink:1998}. These states are shown in Fig. {\ref{fig_1_CH3I_PES}}, which is based on data from Ref. {\onlinecite{alekseyev:2007}}. At the 248nm wavelength ($h\nu$=4.99eV) used in the present work, the $X\rightarrow {^3Q_0}$ excitation dominates gas-phase photoabsorption and the $X\rightarrow {^1Q_1}$ is a minor channel. The subsequent dissociation can proceed via two principal pathways:
\begin{equation} \label{Equ_1}
\begin{split}
CH_3I + h\nu & \rightarrow CH_3 + I(^2P_{3/2}) \textrm{ \{ground state I\}} \\
& \rightarrow CH_3 + I^*(^2P_{1/2}) \textrm{ \{excited I\}}
\end{split}
\end{equation}

The energy difference between ground state I and excited I* is 0.943eV, leading to significant differences in the translational energies imparted to the fragments and which can be resolved in time-of-flight measurements. There are also vibrational and rotational energy partitioning differences for the CH$_3$ photofragments along the various pathways. Another significant feature of this system is that the $X\! -\! {^3Q_0}$ excitation is a parallel transition (requiring a component of the incident $\vec{E}$-field along the C--I bond axis), while the ${^1Q_1}$ and $^3Q_1$ excitations are perpendicular. This polarization dependence for optical absorption allows utilizing polarization and molecular orientation to aid in understanding the photodissociation dynamics at 248nm\cite{Jensen:2005}. As can be seen in Fig. {\ref{fig_1_CH3I_PES}}, the $^3Q_0$ state correlates to the I$^*$ outcome in Equ.~{\ref{Equ_1}}, but the curve-crossing with the $^1Q_1$ state (which correlates to the I pathway) during dissociation enables non-adiabatic transitions that result in both pathways being observed subsequent to an initial $^3Q_0$ excitation\cite{Eppink:1998}. In the gas-phase, photodissociation via the $^3Q_1$ excited state occurs at the long wavelength end of the A-band and proceeds exclusively via the ground state I pathway.

\subsection{Energetics of Stimulated Dissociation}
\label{CH3X_energetics}
The dissociation of a CH$_3$I molecule in free space requires momentum and energy conservation, which determines how the excess kinetic energy is partitioned between the CH$_3$ fragment and the iodine atom. For neutral photodissociation, a starting point for rationalizing the CH$_3$ photofragment kinetic energy is:

%multiple line version
\begin{eqnarray}
T_{CH_3}
& =\frac{m(I)}{m(CH_3I)} \{ &E_{exc} - D_0(C-I) -  E_{int}(I) \nonumber \\
& & - E_{int}(CH_3) \}
\label{Equ_2}
\end{eqnarray}

where $m()$ is the mass of the particular species, $E_{exc}$ is the excitation energy ($h\nu$ for photons), $D_0$ is the C--I bond energy (2.39 eV\cite{Eppink:1998}), $E_{int}$(I) allows for the possible electronic excitation of the departing halogen atom, and $E_{int}$(CH$_3$) is the internal energy (vibration and rotation) of the departing methyl fragment.

For adsorbate systems the parent molecule is not in free space, but embedded at or near the vacuum interface of the system being studied.  It is known from prior work in gas-phase cluster and surface photochemistry that the observed fragment kinetic energy distributions can be altered by chemical or post-dissociation interactions. The departing CH$_3$ fragments must also escape the surface attractive forces, so the kinetic energies will be reduced by an amount on the order of 0.1eV\cite{altaleb:2017}. In any case, Equ.~{\ref{Equ_2}} provides a basis to begin rationalization of the observed kinetic energy distributions.

\section{Experimental Details}
\label{expt_detail}
The experiments were performed in an ultra-high vacuum (UHV) system that has been described previously\cite{Jensen:2005}. The Cu(100) single crystal sample is 12mm in diameter and was cooled by liquid nitrogen (base temperature $\sim$90K) and heated by radiative heating from a filament up to 300K and by electron bombardment heating to 920K for cleaning. Sample temperatures were monitored by a type K thermocouple spot-welded to the tungsten sample mounting wire. Sample cleanliness and order were monitored by Auger electron spectroscopy (AES) and low energy electron diffraction (LEED) measurements respectively. The crystal was prepared in UHV by cycles of Ar$^+$ ion bombardment and electron bombardment heating and annealing until the sample AES spectra indicated a clean copper substrate with the LEED patterns of a (1$\times$1) surface.

Deposition of molecules on the sample was done using a micro-capillary array directed doser\cite{Fisher:2005uw}, with the sample held normal to the doser, 25mm away. This arrangement was found to enhance the deposition by a factor of 10 compared to background dosing. The deposition of the molecules was normally done using substrate temperatures below 100K. The pressure in the UHV chamber was measured using uncorrected ionization gauge readings. The dosing (in Langmuirs, L) was calibrated in terms of equivalent monolayers for the different species used by Temperature Programmed Desorption (TPD) measurements as discussed in Section {\ref{TPD_section}} below. Vapour from CH$_3$I (Sigma-Aldrich, 99.5\%), benzene (Sigma-Aldrich, 99.8\%), fluorobenzene (Sigma-Aldrich, 99\%), 1,4-difluorobenzene (Sigma-Aldrich, \textgreater 99\%), pentafluorobenzene (TCI, \textgreater 98\%), hexafluorobenzene (Sigma-Aldrich, \textgreater 99.5\%), toluene (Sigma-Aldrich, 99.8\%), phenylacetylene (Sigma-Aldrich, 98\%) and CCl$_4$ (carbon tetrachloride, Sigma-Aldrich, \textgreater 99.5\%) was obtained from room-temperature liquid in a pyrex vial a few cm from the precision leak valve used to admit the room-temperature vapour to the directed doser. Vapour from CF$_3$I was admitted to the gas-handling system from a lecture bottle (Sigma-Aldrich, \textgreater 99\%). To deposit lower vapour-pressure aromatics, a PID-controlled heating system was used to raise the temperatures of the pyrex vial and leak-valve assembly up to the doser. This was required for dosing phenol (at 70$^{\circ}$C; Sigma-Aldrich, 99\%), naphthalene (at 80$^{\circ}$C, Sigma-Aldrich, 99\%) and 1-fluoronaphthalene (at 80$^{\circ}$C, Sigma-Aldrich, 99\%). The liquids used in this work were degassed by multiple freeze-pump-thaw cycles and the solids were pumped while cycled between liquid and solid phases to degas. The molecules admitted to the UHV system were checked for impurities using the main chamber mass spectrometer described below.

Temperature programmed desorption (TPD) measurements were performed by positioning the sample to face a quadrupole mass spectrometer (QMS, Stanford Research Systems RGA200). The QMS ionizer was located $\sim$80mm away from the sample and behind an aperture that limits the ionizer line-of-sight to the central region of the sample. The sample was heated using a filament at ground potential, located a few mm behind the sample mount. 

The time-of-flight (TOF) photodissociation measurements were performed using a second QMS (Extrel). Neutral products from the Cu(100) surface travel 185mm to pass through a 4mm diameter aperture to a differentially pumped region with an axial electron bombardment ionizer. The sample to ionizer distance is 203mm. Ions created in the ionizer travel through the quadrupole region and are mass selected, then detected by a conversion dynode and channel electron multiplier (DeTech). Ion arrivals are recorded using a multichannel scaler that begins counting 50$\mu s$ prior to the initiating laser pulse, and the counts recorded from multiple laser pulses are summed. Except where otherwise indicated, the spectra shown in the present work are the result of summing data from 1000 laser pulses into 1000 1$\mu s$ time bins. In order for the ion arrival times to reflect the neutral fragment time-of-flight, they are corrected for the ion flight time $\tau$ in the QMS ($\tau$=4.13$\sqrt{m}$ $\mu s$, for $m$ in amu). This is the leading systematic uncertainty in the recorded flight times ($\pm 1.0\mu s$) which does not affect comparisons between different TOF spectra but does lead to fixed nonlinear systematic uncertainty in the reported fragment kinetic energies $(KE\propto 1/(TOF)^2)$, which is most problematic at short flight times. The TOF spectra $N(t)$ were converted to probability distributions $P(E)$ versus CH$_3$ kinetic energy using the Jacobian transformation with an added factor $1/t$ to account for the higher ionization probability for slower neutral photofragments\cite{zimmermann:1995}. 

The laser pulses ($\sim$5ns duration) are produced by a small excimer laser (MPB PSX-100) operating at 20Hz. In this work KrF ($\lambda$=248nm, $h\nu$=4.99eV) laser light was used, with laser fluences on the sample of $\sim$0.8mJ/cm$^2$. The intrinsic bandwidth of the laser emission for excimer lasers is rather broad-- for a free-running KrF excimer laser the center wavelength is approximately 248.4nm (4.992eV) and has a fwhm bandwidth of $\sim$0.40nm (0.008eV). 

Linearly polarized laser light has been used exclusively in this work for the reasons described in Section {\ref{CH3X_pdissn}}. To create the polarized light, the beam is incident upon a birefringent MgF$_2$ crystal prism at the Brewster angle to separate the p- and s-polarized components. The p-polarized beam was aligned on the sample. The s-polarized light was derived from the p-polarized beam by inserting an antireflection coated zero-order half-waveplate into the beam. The laser light was collimated using a 6mm diameter aperture and was unfocused on the sample. The laser light is incident upon the sample at a fixed angle of 45$^\circ$ from the TOF mass spectrometer axis-- for example, when the Cu crystal sample is oriented to collect desorption fragments along the surface normal direction, the light is incident at 45$^\circ$.

Cross sections for the photodissociation of molecular thin films examined in this work were determined by measuring the depletion rate of the CH$_3$ or CF$_3$ photofragment yields (summed photofragment counts). These ``depletion cross sections'' are obtained by recording photofragment yields from photodissociation for a sequence of TOF spectra. Time-of-flight spectra are obtained using 100--400 laser pulses per scan, then repeated for 10 or more successive scans. In the systems studied here, the photofragment yields are observed to monotonically diminish as the net laser photon flux was increased, and the resulting yield vs. photon flux curves were then fit using a single exponential decay model. Reasonable fits to the data were obtained, in the low photon flux region. This procedure allows the possibility that other photochemical processes involving CH$_3$I or CF$_3$I removal but not seen in the TOF data might be occurring in the heterogeneous thin films. In the present work, the dissociated species was always from a dose equivalent to 1ML CH$_3$I or CF$_3$I film, while the underlying aromatic thin film was varied. We used 1ML CH$_3$I or CF$_3$I on a 10ML hexafluorobenzene thin film as a reference system for comparison when studying the cross sections of the respective thin film systems. The advantage of this approach is that the errors in comparative relative cross sections (e.g. vs CH$_3$I on hexafluorobenzene) between the different systems is low, on the order of $\sim$10\% based on repeatability of measurements, as compared to determination of the absolute cross section, where we expect errors on the order of $\pm$50\%.

\section{Observations}
\subsection {Temperature Programmed Desorption}
\label{TPD_section}
In order to be able to adsorb known quantities of the various molecules used to create thin films in the present work, TPD has been used to characterize the relationship between the amounts dosed (in Langmuirs, L) and the formation of a complete monolayer (1ML). After establishing the dose required to form 1ML for a particular molecule, we have assumed that for the temperature used in these experiments (T\textless 100K) the sticking coefficient for these molecules is close to unity for multilayer films, so dose amounts in multiples of the monolayer dose to produce the desired multilayer thin film. We have characterized TPD for several of the molecules used in the present work (CH$_3$I, fluorobenzene, 1,4-difluorobenzene and pentafluorobenzene) in our apparatus previously{\cite{jensen:2024}} and have performed similar TPD experiments to establish monolayer doses the other molecules used in the present work. Characterizing the adsorption of phenylacetylene on Cu(100) was more difficult as the alkyne group forms a stronger bond on copper surfaces such that on Cu(111) the monolayer desorbs at 410K\cite{sohn:2007}-- we found that on Cu(100) the monolayer desorption temperature was above the upper limit of our TPD system. To establish the dose required for the phenylacetylene monolayer, we created a Cl-terminated Cu(100) surface by adsorbing a monolayer of CCl$_4$ at 100K and warming to 350$^{\circ}$C (repeated twice), a procedure that allows thermal dissociation of the CCl$_4$ and desorption of the chlorocarbon products to leave a Cu(100)-Cl surface. We found that this substrate is more inert and allowed the phenylacetylene to be adsorbed and desorbed intact (detected at 102amu by the QMS), with the monolayer desorption peak at 215K while the second and subsequent layers desorb with a peak at 190K. For the photochemical measurements using phenylacetylene films, we generally used the Cu(100)-Cl substrate to avoid the buildup of surface carbon when the substrate was cleaned by heating between scans.

\subsection{Photodissociation of CH$_3$I on Benzene and Fluorobenzene Thin Films}
\label{fluorobenz}
\begin{figure}[t]
\includegraphics[scale=0.65]{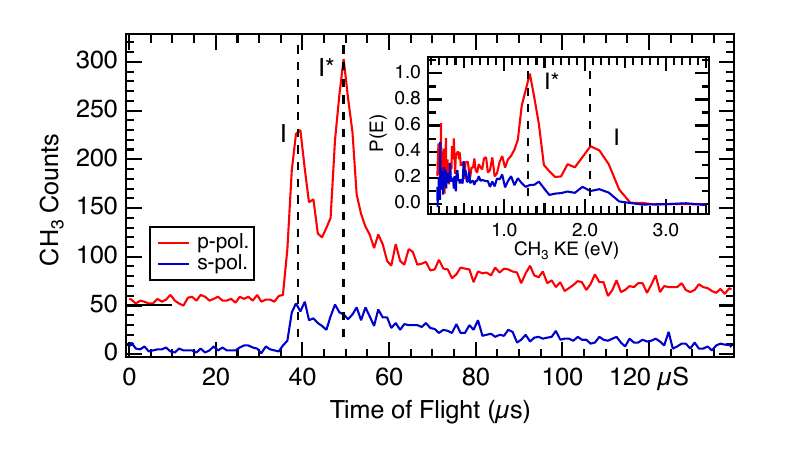}
\caption{Time-of-flight spectra for CH$_3$ photofragments from 1ML CH$_3$I on a 5ML thin film of pentafluorobenzene, obtained using 248nm light. The light is incident at 45$^{\circ}$ from the surface normal and the CH$_3$ photofragments are detected in the surface normal direction. The inset plot shows the CH$_3$ probability distribution plotted as a function of the CH$_3$ fragment kinetic energy. The vertical dashed lines show the nominal peak flight times and energies respectively for the I and I* dissociation pathways described in the text (Section {\ref{CH3X_pdissn}}). }
\label{fig_CH3I_on_pFBz}
\end{figure}

A time-of-flight spectrum obtained using 248nm light from 1ML of CH$_3$I on a 5ML thin film of pentafluorobenzene is shown in Fig. {\ref{fig_CH3I_on_pFBz}}. Using p-polarized light, distinct peaks from CH$_3$ photofragments collected in the surface normal direction are observed at flight times of 39$\mu$s and 49$\mu$s. These features correspond to the I and I* dissociation channels respectively, as described in Section \ref{CH3X_pdissn}. Using incident p-polarized light, there is a component of the $\vec{E}$ field perpendicular to the surface, so that molecules with the C--I bond oriented in the surface normal direction can absorb light via the transition to the $^3Q_0$ state. For the p-polarized spectrum, a substantial fraction of the molecules initially excited to the $^3Q_0$ state cross over to the $^1Q_1$ state during dissociation (as seen in Fig. {\ref{fig_1_CH3I_PES}}), leading to photofragments in the I channel. Using s-polarized light ($\vec{E}$ in the surface plane), the overall signal in Fig. {\ref{fig_CH3I_on_pFBz}} is substantially reduced but a small CH$_3$ photofragment peak in the I channel is observed, while there is no peak seen for the I* channel. The explanation for the observed s-polarized TOF spectrum is that photoabsorption occurs via the $^1Q_1$ excited state\cite{gardiner:2015} (a perpendicular transition) to produce the peak at 39$\mu$s flight time, and that some inelastic scattering for CH$_3$ produced in the surface region creates the continuum of CH$_3$ flight times observed. In summary, the features of Fig. {\ref{fig_CH3I_on_pFBz}} can be understood based on the gas-phase model for CH$_3$I photodissociation described in Section \ref{CH3X_pdissn} and Fig. {\ref{fig_1_CH3I_PES}}. The sensitivity of the CH$_3$I neutral photodissociation to the incident laser light polarization is used as a tool in understanding the EET photodissociation mechanism we observe in other systems we describe in this work.

\begin{figure}[b]
\includegraphics[scale=0.75]{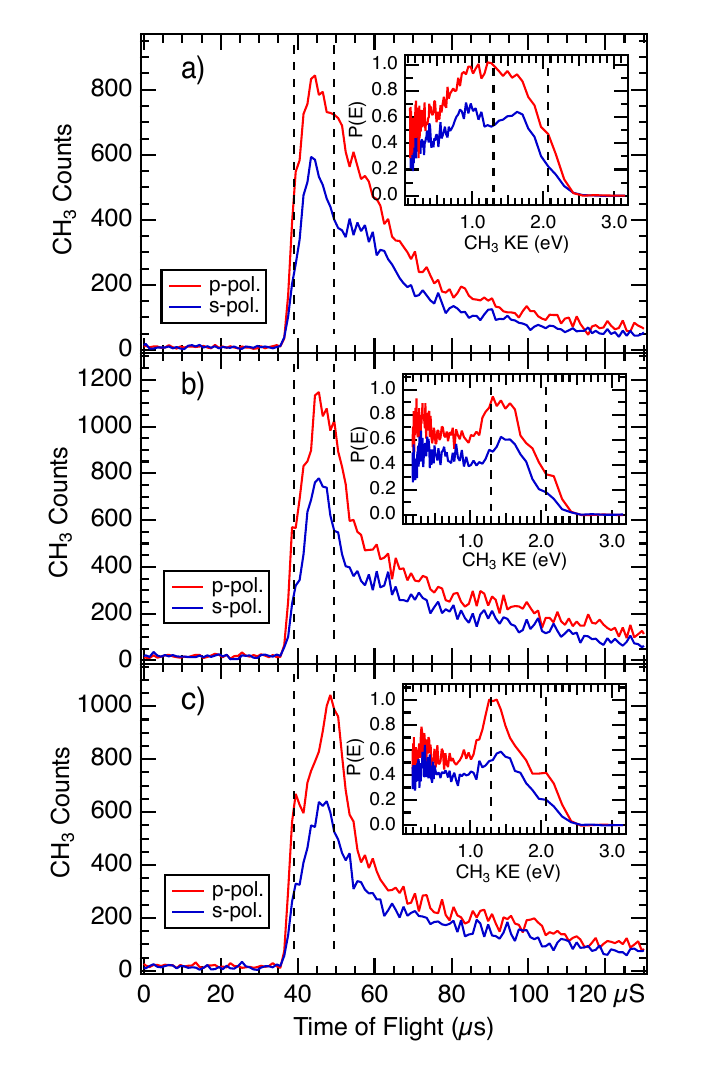}
\caption{Time-of-flight spectra for CH$_3$ photofragments from 1ML CH$_3$I on 10ML thin films of: (a) benzene; (b) fluorobenzene; and (c) 1,4-difluorobenzene. The inset plots show the data as probability distributions as a function of the CH$_3$ kinetic energy. The vertical dashed lines indicate the nominal I and I* neutral photodissociation pathway peak times or energies in the plots. }
\label{fig_CH3I_on_mFBz_dFBz}
\end{figure}

Time-of-flight spectra from 1ML CH$_3$I adsorbed on 10ML thin films of benzene, fluorobenzene and 1,4-difluorobenzene are shown in Fig. {\ref{fig_CH3I_on_mFBz_dFBz}}. In contrast with the spectra of Fig. {\ref{fig_CH3I_on_pFBz}}, the s-polarization TOF spectra show a significant dissociation pathway, with a peak intermediate to the I and I* channels observed for neutral CH$_3$I photodissociation: at flight times of 44$\mu$s, 46$\mu$s and 47$\mu$s for benzene, fluorobenzene and difluorobenzene respectively. The same new dissociation feature contributes to the p-polarization TOF spectra in Fig. {\ref{fig_CH3I_on_mFBz_dFBz}}, seen in addition to the I and I* channel peaks. As a consequence of the observed additional dissociation pathway in the TOF spectra, the depletion cross sections measured for CH$_3$I on thin films of benzene, fluorobenzene and 1,4-difluorobenzene are increased (by factors of 2.0$\times$, 3$\times$ and 1.8$\times$ respectively) relative to that for CH$_3$I on hexafluorobenzene\footnote{The observed depletion cross sections depend on the thickness of the thin film used (e.g. Section {\ref{phen_depletion}})-- we have used 10ML films as a reference for comparisons.}. In contrast, the depletion cross section for CH$_3$I on pentafluorobenzene (such as in Fig. {\ref{fig_CH3I_on_pFBz}}) is the same as for the reference hexafluorobenzene system, consistent with the TOF spectra showing only the gas-phase-like CH$_3$I photodissociation\cite{jensen:2024}.

The observations made from CH$_3$I photodissociation on thin films of benzene and a range of fluorobenzenes has been discussed  previously\cite{jensen:2024}. The new dissociation pathway observed in the p- and s-polarization TOF spectra is due to photoabsorption in the thin film by electronic excitation of the aromatic molecules, with subsequent energy transfer (and also likely energy transport within the thin film) to the CH$_3$I adsorbed on top. This EET mechanism decouples the dissociation energy from the photon energy, and instead is determined by the electronic excitation transferred to the CH$_3$I. That this mechanism is not observed for the higher fluorobenzenes (e.g. pentafluorobenzene as in Fig. {\ref{fig_CH3I_on_pFBz}}) is likely due to the much shorter lifetimes for the excited intermediate states for these molecules. Another observation made for this effect is that the magnitude of the enhancement does not simply correlate with the initial absorption cross section of the molecules in the thin film at 248nm-- for example the gas-phase absorption cross section for 1,4-difluorobenzene is $\sim$50\% greater than that for fluorobenzene, but the measured depletion cross sections show an opposite trend. For the adsorbed multilayer system, the depletion cross sections also depend on steps subsequent to the initial photoabsorption, including the excitation lifetime, the probability of energy transfer to co-adsorbed CH$_3$I, and possibly also the mobility of the excitation within the aromatic thin film.

\subsection{Photodissociation of CH$_3$I on Substituted Benzenes}
\label{monobenz_section}
\begin{figure}[t]
\includegraphics[scale=0.70]{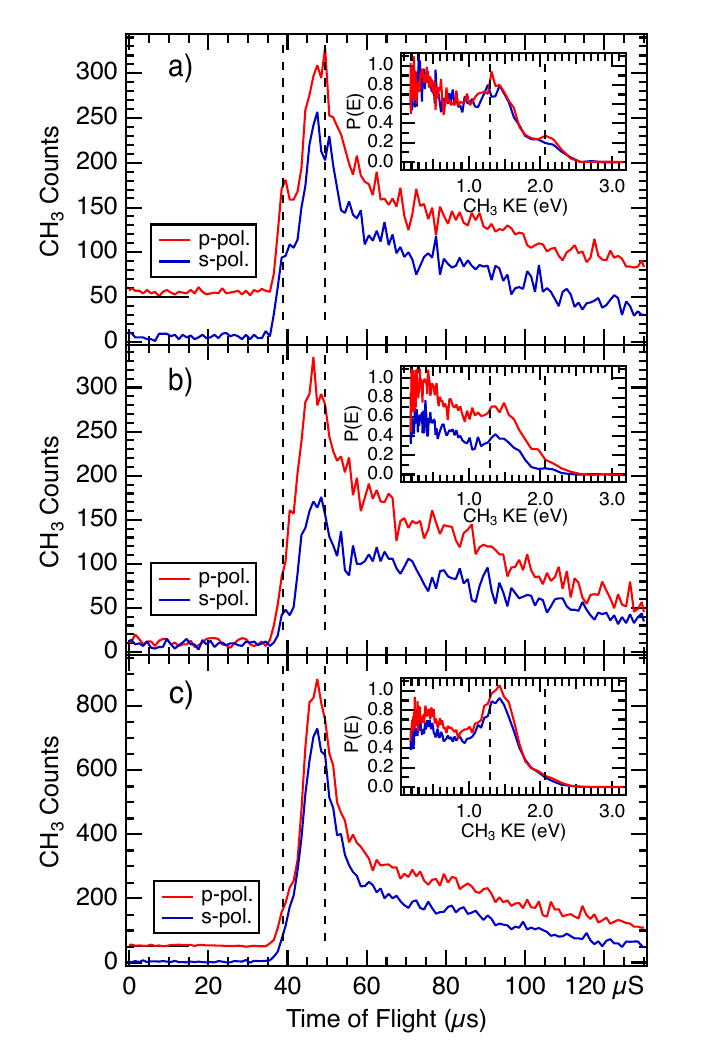}
\caption{Time-of-flight spectra for CH$_3$ photofragments from 1ML CH$_3$I on 10ML thin films of: (a) phenol; (b) toluene; and (c) phenylacetylene. The p-polarization plots for (a) and (c) have been offset vertically by 50 counts to separate the spectra for clarity. The data for (a) and (b) are obtained  from 1000 laser pulses while that for phenylacetylene in (c) is obtained from 200 laser pulses and using a Cu(100)-Cl surface. }
\label{fig_CH3I_on_C6H5X}
\end{figure}

Time-of-flight spectra for CH$_3$ photofragments from 1ML CH$_3$I on 10ML thin films of phenol, toluene and phenylacetylene are shown in Fig. {\ref{fig_CH3I_on_C6H5X}}. The observed features in these TOF spectra are qualitatively similar to those of the fluorobenzenes in Fig. {\ref{fig_CH3I_on_mFBz_dFBz}}, with a significant dissociation pathway observed for both s- and p-polarized light in addition to the I and I* CH$_3$I photodissociation pathways seen predominantly for p-polarized light. Both the p- and s-polarization TOF spectra of Fig. {\ref{fig_CH3I_on_C6H5X}} (a)--(c) display the EET dissociation peak at an intermediate TOF that peaks at a similar flight time of 48$\mu$s (1.43eV) for these aromatic thin films. The CH$_3$ photofragment energies observed for the EET feature for the all the mono-substituted benzenes and difluorobenzene fall in a narrow range, albeit slightly higher energy for CH$_3$I on the fluorobenzene thin films. The CH$_3$I neutral photodissociation I and I* TOF features do not appear to vary for the different thin films used, however the position and magnitude of the overlapping I* feature is difficult to determine precisely in most of these spectra. It is not clear if the CH$_3$ photofragment kinetic energy differences for the EET feature arise from variation in the excitation energy transferred to the CH$_3$I from the thin film or reflect differing morphologies at the thin film interface that change the binding site for the CH$_3$I. As discussed previously for CH$_3$I on fluorobenzene, the thin film morphology is likely to be complex for these multilayer aromatic films\cite{jensen:2024}. One consequence of this could be the proportions of the CH$_3$ photofragment signal that is present in the well-defined TOF peaks compared to that in the inelastic continuum and tail of the spectra (e.g. signal at $\tau$\textgreater 60$\mu$s). For example, the inelastic continuum is more pronounced for CH$_3$I on phenol and toluene films (Fig. {\ref{fig_CH3I_on_C6H5X}} (a) \& (b)) than that for fluorobenzene, difluorobenzene and phenylacetylene thin films.  

The systems shown in Fig. {\ref{fig_CH3I_on_C6H5X}} display enhanced CH$_3$I depletion cross sections compared to the reference system-- for phenol and toluene, the depletion cross sections for 10ML thin films are similar to those measured for fluorobenzene and difluorobenzene (2--3$\times$ that of the reference system). Similar to the discussion in Section {\ref{fluorobenz}}, there is no direct correlation between the gas-phase 248nm photoabsorption cross section for the thin film aromatic and the observed depletion cross sections. For example, at 248nm the gas-phase phenol absorption cross section is 4$\times$ larger than that of toluene, but the depletion cross sections of CH$_3$I on these thin films are essentially the same. 

The depletion cross sections measured for CH$_3$I on phenylacetylene thin films are much different-- approximately 20$\times$ larger than that of the reference system. The CH$_3$ yield and cross section is so large that the number of laser pulses used for TOF spectra (such as the data in Fig. {\ref{fig_CH3I_on_C6H5X}}(c)) was limited to 200 pulses so that the spectra reflect dissociation of the initial CH$_3$I present in the monolayer\footnote{For the CH$_3$I on phenylacetylene system, the TOF spectra from the depleted system (e.g. 80\% signal reduction) were not significantly different from the initial spectra aside from the lower signal rate.}.  The TOF spectra in Fig. {\ref{fig_CH3I_on_C6H5X}}(c) are nearly identical for the p- and s-polarized light illumination, in contrast to the other substituted benzenes thin films studied. The EET mechanism for CH$_3$I dissociation on the 10ML phenylacetylene thin film dominates the neutral CH$_3$I photodissociation channels even when p-polarized light is used. For the phenylacetylene system the initial photoabsorption cross section at 248nm is clearly a significant factor in the magnitude of the CH$_3$I depletion cross section we observe, again in contrast to the other substituted benzenes studied here. While we are not aware of gas- or condensed-phase photoabsorption cross section measurements for phenylacetylene, in {\it{n}}-heptane solution\cite{yamakawa:1968} it is found to be $\sim$100$\times$ larger than that of benzene at 248nm, as compared to the 1--5$\times$ factors for the other substituted benzenes studied in the present work. The larger cross section for phenylacetylene appears to be a consequence of the second singlet excited state (S$_2$), $^1\!A_1$,  being lower in energy than for the other substituted benzenes, somewhat more so for condensed-phase phenylacetylene\cite{swiderek:1998}, such that 248nm light ($h\nu$=4.99eV) is just above the onset energy for this absorption band. Due to the large photoabsorption in the phenylacetylene thin film, the EET mechanism for dissociation of CH$_3$I dominates the TOF spectra (Fig. {\ref{fig_CH3I_on_C6H5X}} (c)) to the extent that using p-polarized light to access neutral CH$_3$I photodissociation, the added intensities from I and I* channels are not significant features in the TOF spectrum. For the reasons outlined in Section {\ref{TPD_section}} most of the data collected for CH$_3$I on phenylacetylene thin films used the Cu(100)-Cl as a substrate\footnote{Due to the higher binding energy for phenylacetylene on Cu(100), radiative heating alone did not remove the first layer, and electron bombardment heating of the substrate caused dissociation of these molecules leading to the accumulation of surface carbon. The weaker binding for phenylacetylene on the Cu(100)-Cl substrate circumvented this issue.}. Some spectra and measurements were also obtained by dosing the phenylacetylene thin film on the Cu(100) surface--  no observable differences in the TOF spectra due to the substrate used were found.

\subsection{Photodissociation of CH$_3$I on Naphthalenes}
\label{naph_section}
\begin{figure}[b]
\includegraphics[scale=0.70]{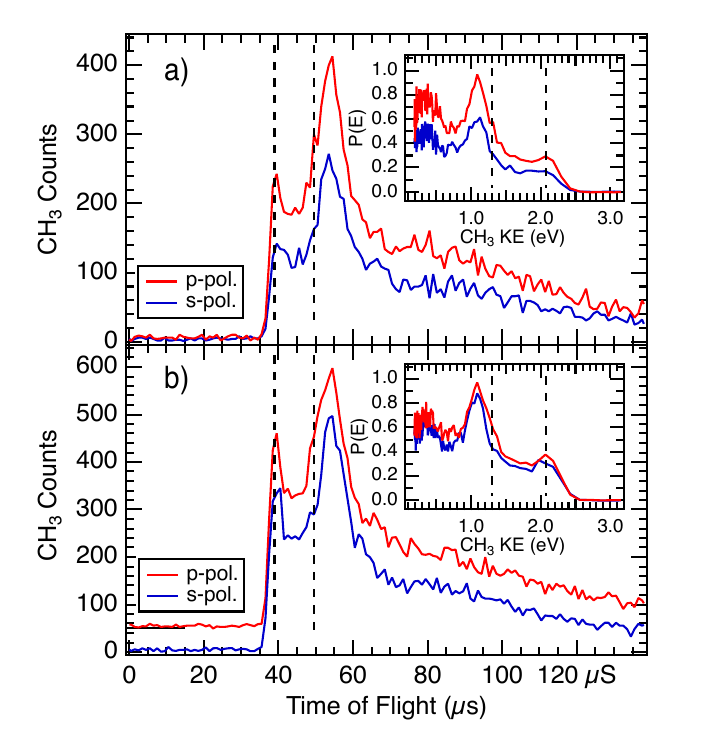}
\caption{Time of flight spectra for CH$_3$ photofragments from 1ML CH$_3$I on 10ML thin films of: (a) naphthalene; and (b) 1-fluoronaphthalene. The p-polarization data in (b) is offset vertically by 50 counts to separate the spectra for clarity. }
\label{fig_CH3I_on_naphs}
\end{figure}

Figure {\ref{fig_CH3I_on_naphs}} shows CH$_3$ photofragment TOF spectra from 1ML CH$_3$I adsorbed on 10ML thin films of naphthalene and 1-fluoronaphthalene. The most prominent feature of these spectra is the peak at 54$\mu s$ flight time (1.12eV kinetic energy) seen for both p- and s-polarized 248nm light. This peak appears to be a consequence the EET mechanism identified for the substituted benzenes (Figs. {\ref{fig_CH3I_on_mFBz_dFBz}} and {\ref{fig_CH3I_on_C6H5X}}) but observed at longer flight times. The lower CH$_3$ photofragment kinetic energies for the EET mechanism in these systems is interpreted as being due to a smaller excitation energy being imparted to the CH$_3$I from the thin films of the naphthalenes. 

The TOF spectra of Fig. {\ref{fig_CH3I_on_naphs}} also show features from the I and I* pathways of 248nm CH$_3$I photodissociation (as per Section {\ref{CH3X_pdissn}})-- the p-polarization spectra display a small shoulder at 50$\mu s$ (I* features) that is absent in the corresponding s-polarization spectra. The I-pathway peak seen at 39$\mu s$ flight time is quite prominent in all of the spectra of Fig. {\ref{fig_CH3I_on_naphs}}, notably more so than for the substituted benzene thin film systems (Figs. {\ref{fig_CH3I_on_mFBz_dFBz}} and {\ref{fig_CH3I_on_C6H5X}})-- for CH$_3$I on the naphthalene thin films this feature is observed in both the p- and s-polarization spectra. As discussed in Section {\ref{CH3X_pdissn}}, the I-pathway is allowed for s-polarized light via a perpendicular transition to the $^1Q_1$ state, where little or no dissociation via the I* pathway is anticipated\footnote{For gas-phase CH$_3$I the curve-crossing between the $^3Q_0$ and $^1Q_1$ states is understood to be more effective in the ${^3Q_0} \rightarrow {^1Q_1}$ direction than the converse due to the symmetries in dissociation along the different pathways\cite{townsend:2004}}. In the p-polarization spectra for most of the thin films studied, the relative intensities of the I and I* TOF features are difficult to compare due to the I* feature overlap with the EET dissociation feature-- for example, in Fig. {\ref{fig_CH3I_on_pFBz}} the I* peak is seen to be larger than the I peak. In our experimental geometry (CH$_3$ detection in the surface normal direction, light incident at 45$^{\circ}$), incident p-polarized light has components perpendicular and parallel to the surface, while s-polarized light is purely parallel to the surface plane. The apparently enhanced I-pathway signal seen using s-polarized light for CH$_3$I on the naphthalene thin films using s-polarized light is intriguing. The same effect might also be present for dissociation on the substituted benzene films (Figs. {\ref{fig_CH3I_on_mFBz_dFBz}} and {\ref{fig_CH3I_on_C6H5X}}) but less evident due to the stronger overlap with the large EET feature at shorter flight times in those spectra. One possible explanation is that this effect is due to symmetry breaking for CH$_3$I in the adsorbed state for these thin films, leading to an enhanced contribution from either the $^1Q_1$ or $^3Q_1$ states, analogous to a suggestion made for (CH$_3$I)$_2$ dimer photodissociation\cite{denalda:2011}.

Depletion cross sections for the CH$_3$I on 10ML naphthalene and 1-fluoronaphthalene thin films found an enhancement of $\sim$2.5$\times$ as compared to the reference system, similar in magnitude to that found for the substituted benzenes thin films aside from phenylacetylene. Unlike the substituted benzenes, the $S_1$ excited state for naphthalenes is at an energy well below that of the incident photons. In the region of the 248nm wavelength used, the photoabsorption for naphthalenes  is dominated by the $S_2$ excitation, which in the gas-phase has a cross section $\sim$10$\times$ that of the substituted benzenes\cite{KellerRudek:2013wf}. This is discussed further in Section {\ref{additionaldisc}}.

\subsection{Photodissociation of CF$_3$I on Aromatic Thin Films}
\label{CF3I_pdissn}
The EET dissociation mechanism is also observed for CF$_3$I adsorbed on thin films of aromatic molecules. In the gas-phase, CF$_3$I has an A-band excited state structure very similar to that of CH$_3$I (e.g. Fig. \ref{fig_1_CH3I_PES})\cite{alekseyev:2013}, with a photodissociation cross section at 248nm about one-third that of CH$_3$I\cite{KellerRudek:2013wf}. Photodissociation of CF$_3$I and detection of the CF$_3$ or I/I* photofragments has been studied in the gas-phase\cite{vanveen:1985,furlan:1996,wang:2003,lin:2016} as well as in the adsorbed state\cite{sun:1995}. The CF$_3$I dissociation dynamics are broadly similar to that of CH$_3$I (Section {\ref{CH3X_pdissn}}), with a slightly smaller C--I bond energy ($D_0$=2.32eV) and a larger fraction of the excess energy disposed in internal excitation of the CF$_3$ photofragments\cite{furlan:1996,lin:2016}. The higher mass of the CF$_3$ photofragment leads to it having lower speeds and kinetic energies as compared to the CH$_3$/CH$_3$I case as per Equ. {\ref{Equ_2}}.

Several previous studies have been made of CF$_3$I adsorbed on coinage metal surfaces\cite{castro:1993,szabo:1996}. Structural differences for CF$_3$I as compared to CH$_3$I adsorbed on the molecular thin films in the present work are likely to arise from steric and electrostatic differences-- for example CF$_3$I and CH$_3$I have similar dipole moments (1.048D and 1.6406D respectively)\cite{CRC_Handbook:2024} but in opposite directions, as the CF$_3$ moiety has a net negative partial charge\cite{cox:1980}. 

Detection of the CF$_3$ photofragments in the QMS electron ionizer was found to have significantly higher count rates for the CF$_2^+$ ion fragment than the CF$_3^+$ or CF$^+$ ions\cite{sun:1995}, and to the extent we studied, the TOF profiles were the same after correcting for the differing ion transit times in the QMS. Accordingly, the CF$_3$ TOF spectra we present were obtained measuring the CF$_2^+$ (50amu) ion fragments. 

\begin{figure}[t]
\includegraphics[scale=0.80]{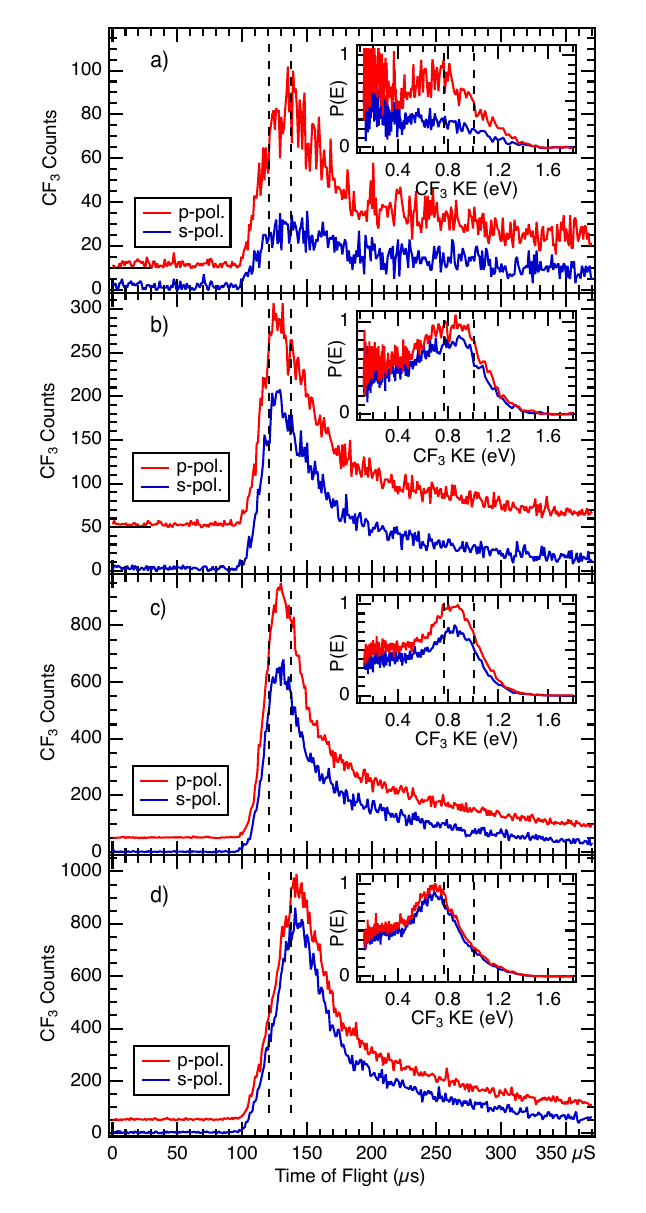}
\caption{Time of flight spectra for CF$_3$ photofragments from 1ML CF$_3$I on 10ML thin films of (a) hexafluorobenzene; (b) fluorobenzene; (c) phenylacetylene; and (d) 1-fluoronaphthalene, obtained using 248nm light. The vertical dashed lines indicate the nominal I and I* neutral photodissociation pathway peak times or energies, as described in the text. The spectra are obtained using 1000 laser pulses except for (c) where 400 pulses were used.}
\label{fig_CF3I_on_various}
\end{figure}

Figure {\ref{fig_CF3I_on_various}} shows the CF$_3$ photofragment TOF spectra obtained from monolayer quantities of CF$_3$I adsorbed on various aromatic thin films. In Fig. {\ref{fig_CF3I_on_various}}(a), the CF$_3$ TOF spectra from CF$_3$I adsorbed on 10ML hexafluorobenzene displays a fairly low signal rate that is similar to the findings for CH$_3$I on 10ML hexafluorobenzene or pentafluorobenzene films-- with a dissociation cross section close to the gas-phase value. It is notable that unlike in the gas-phase study\cite{vanveen:1985}, the I/I* pathways do not lead to CF$_3$ photofragment peaks in the TOF spectra that can be visually separated. Gas-phase photodissociation of CF$_3$I at 248nm leads predominantly to the I* channel-- the $\Phi^*$ ratio ($\Phi^*=\frac{N(I^*)}{N(I) + N(I^*)}$) is large (0.92)\cite{vanveen:1985}. In the adsorbed state $\Phi^*$ is smaller\cite{sun:1995}, similar to the trend observed for CH$_3$I between the gas-phase and adsorbed state (e.g. Fig. {\ref{fig_CH3I_on_pFBz}}). One effect of the reduced $\Phi^*$ seems to be peaks of similar magnitude in the I and I* channels, so overlapping of the CF$_3$ photofragment peaks from these is  stronger than in the gas-phase. Although our TOF path length is only somewhat less than that in the gas-phase study of Ref. \onlinecite{vanveen:1985} (203mm vs 235mm), the combination of factors of the path length, lowered $\Phi^*$, and broader CF$_3$ TOF peak profiles (from CF$_3$ internal excitations\cite{vanveen:1985,furlan:1996,wang:2003} and possibly the surface environment) has made separating the I and I* pathways more difficult. The vertical dashed lines in Fig. {\ref{fig_CF3I_on_various}} are located based on scaling the gas-phase TOF data\cite{vanveen:1985} to show the location of the I and I* pathway peaks observed in that work, intended to serve as references for the variations found in our TOF data. In Fig. {\ref{fig_CF3I_on_various}}(a), switching between p- and s-polarized incident light should have the same effect for CF$_3$I as found for CH$_3$I, suppressing the dominant $^3Q_0$ excitation pathway for molecules with the C--I bond axis in the surface normal direction by using s-polarized light. A reduced CF$_3$ signal is observed, but not the unambiguous switching such as that found for CH$_3$I (e.g. as in Fig. {\ref{fig_CH3I_on_pFBz}}). A variety of other CF$_3$I adsorbate systems were prepared and studied to see if the I/I* pathways could be more clearly separated in other circumstances, but so far we have found none.

For CF$_3$I adsorbed on 10ML fluorobenzene (Fig. {\ref{fig_CF3I_on_various}}(b)), the CF$_3$ TOF spectra shows a significantly increased yield with peaks in the p- and s-polarized spectra at $\sim$127$\mu s$. The differences observed between the CF$_3$ photofragment spectra in Figs {\ref{fig_CF3I_on_various}} (a) and (b) are similar to that seen for CH$_3$I on the same thin films-- increased yields and corresponding depletion cross section (by $\sim 3\times$ relative to that of Fig. {\ref{fig_CF3I_on_various}}(a)) on the fluorobenzene thin film as well as similar yields for both light polarizations, with the TOF peaks intermediate between the I and I* pathway peaks. The TOF spectra from 1ML CF$_3$I adsorbed on 10ML phenylacetylene (Fig. {\ref{fig_CF3I_on_various}} (c)) display the largest yield and depletion cross sections for all of the aromatics studied (for CF$_3$I, enhanced by $10\times$ compared that on the reference C$_6$F$_6$ thin film), very similar to what is found in the CH$_3$I case. The peaks in the p- and s-polarization spectra are located similarly to that from the fluorobenzene thin films, at $\sim$129$\mu s$, indicating that similar dissociation dynamics underly both systems, with the excitation causing dissociation lower in energy than the incident photons. This interpretation is supported by the observations for 1ML CF$_3$I on 10ML films of 1-fluoronaphthalene (Fig. {\ref{fig_CF3I_on_various}}(d)). Here the CF$_3$ photofragment yield and depletion cross sections are enhanced (by $3\times$). while the CF$_3$ peak arrival time is significantly later, at 142$\mu s$. This is consistent with the trend seen for CH$_3$I on the naphthalene films in Section {\ref{naph_section}}-- an enhanced photoabsorption mediated by the thin film, with a smaller excitation energy transferred to the CF$_3$I that leads to lower kinetic energy imparted to the photofragments.

 \subsection{Substrate Quenching of Photodissociation in Aromatic Thin Films}
 \label{phen_depletion}
\begin{figure}[t]
\includegraphics[scale=0.80]{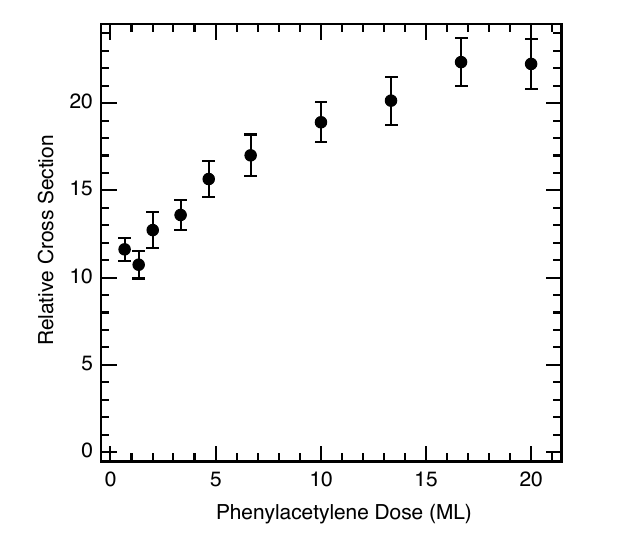}
\caption{Depletion cross sections for 1ML CH$_3$I on varying doses of phenylacetylene on Cu(100)-Cl. The TOF data was obtained using s-polarized light. The vertical scale is relative to that for CH$_3$I on 10ML C$_6$F$_6$ as described in the text. The error bars shown are taken from the standard error from the exponential fitting function. }
\label{fig_Cross_Sec}
\end{figure}

The dissociation of CH$_3$I adsorbed on a thin aromatic film due to photoexcitation within the film competes with the excitation energy being dissipated by various processes within the thin film (e.g. internal conversion or intersystem crossing) or by transfer of the excitation to the substrate. Substrate quenching can reduce the yield of CH$_3$ photofragments per incident photon-- in our previous study, the CH$_3$ photofragment yield from fluorobenzene thin films was very low for films that were a few monolayers thick\cite{jensen:2024}, particularly for the EET mechanism. The very large effective cross section for CH$_3$I dissociation on phenylacetylene thin films is found to support a significant yield even for monolayer films. Depletion cross sections for a range of phenylacetylene doses are shown in Fig. {\ref{fig_Cross_Sec}}. The effective cross section increases by a factor of two from the thinnest 1--2ML films to the thick multilayer films, indicating that the initial photoabsorption can be quenched by proximity to the Cu(100)-Cl substrate. Across the range of phenylacetylene film thicknesses, the form of the TOF spectra are not observed to change from that seen in Fig. \ref{fig_CH3I_on_C6H5X}(c) aside from changing magnitude. While CuCl in bulk form is a wide gap semiconductor, the electronic structure of the Cu(100)-Cl interface is less well characterized than its physical structure\cite{andryushechkin:2018}. We have found previously that CH$_3$I photodissociation is quenched for molecules in direct contact with a halogen passivated Cu surface (Cu(110)-I)\cite{johnson:2000} but the second layer of CH$_3$I molecules was photodissociated efficiently. In that context it is significant that the phenylacetylene photoexcitation for the 1--2ML films is sufficiently long-lived to dissociate the CH$_3$I adsorbed on top. Phenylacetylene adsorbed on Cu(111) was found to be bound via the alkyne group so that the phenyl ring is fairly upright on the surface\cite{qi:2019}. If a similar binding structure is present on the Cu(100)-Cl substrate, the combination of the lowered quenching rate from the chlorinated surface and the photoactive phenyl group suspended above the substrate might account for the substantial photoactivity we observe for CH$_3$I adsorbed on the monolayer phenylacetylene films. It is most likely that for the thinnest films, resonant electron transfer with the substrate\cite{avouris:1984} is the dominant quenching mechanism, similar to the observations of the quenching of luminescence in ultrathin organic thin films grown on metal substrates\cite{gebauer:2004}.

\section{Additional Discussion}
\label{additionaldisc}
In the region of the 248nm wavelength used in the present study, the photoabsorption for the substituted benzenes is generally via $S_1$ states similar to the ${^1\!B_{2u}}$ excited state in benzene. There is limited data available on cross sections of these species in the condensed state-- data from gas-phase\cite{KellerRudek:2013wf} and solution-phase\cite{taniguchi:2018} measurements show values that vary in a limited range (1$\times$--3$\times$ that of benzene). The exception to this trend is phenylacetylene, having a solution-phase cross section at 248nm that is $\sim$100$\times$ larger than that of benzene\cite{yamakawa:1968}. This is due to the onset of the $S_2$ (${^1\!A_1}$) absorption in this energy region, which is also observed in inelastic electron scattering from condensed-phase phenylacetylene\cite{swiderek:1998}. In phenylacetylene, this initial $S_2$ absorption undergoes rapid internal conversion (in the gas-phase, in $\sim$54fs\cite{lee:2002}) to the longer-lived $S_1$ excited state, with a lifetime of $\sim$54ns in the gas-phase\cite{hui:1974}. As a consequence, for CH$_3$I dissociation on phenylacetylene thin films (Fig. {\ref{fig_CH3I_on_C6H5X}}(c)), the CH$_3$ photofragment TOF and kinetic energies are characteristic of excitation via an $S_1$ excited state similar to the other substituted benzenes in Fig. {\ref{fig_CH3I_on_C6H5X}}, but with a much higher effective cross section due to the $S_2$ photoabsorption.

The EET dissociation of CH$_3$I by the thin films of the naphthalenes follows energetics similar to that for phenylacetylene. Illumination with 248nm light results in an initial excitation in the $S_2$ (${^1\!B_{1u}}$) band\cite{watts:1966} of naphthalene, having a large photoabsorption cross section ($\sim\!20\times$ that of the  benzene), which then undergoes a rapid internal conversion to the $S_1$ state manifold (in the gas-phase, in \textless 100fs\cite{schmitt:2001}), so that energy transfer to the CH$_3$I coadsorbate will occur via the long-lived but lower energy $S_1$ excited state. For crystalline naphthalene the $S_1$ band edge is at 3.90eV and the fluorescence lifetime is $\sim$100ns\cite{auweter:1979}. The mean free path for excitons in naphthalene are estimated to be tens of lattice spacings\cite{stehr:2014}, so in principle could traverse the thin films used here. Less detail is available for 1-fluoronaphthalene though the excited state energies are very similar\cite{singh:1981}. 

The ranges of $S_1$ excitation energies and fluorescence bands for the substituted benzenes and naphthalenes are illustrated in Fig. {\ref{fig_1_CH3I_PES}}. Neutral photodissociation of CH$_3$I at 248nm proceeds primarily via the pathways described in Section {\ref{CH3X_pdissn}}-- 248nm p-polarized light results in excitation via the $^3Q_0$ state, while s-polarized light causes excitation of the $^1Q_1$ state. The EET excitations of CH$_3$I from the substituted benzenes and naphthalenes apparently lead exclusively to dissociation via the $^3Q_1$ pathway\footnote{the evidence that EET for CH$_3$I on benzene might proceed in whole or in part via the $^3Q_0$ excitation has been discussed previously\cite{jensen:2021,jensen:2024}.}, as the CH$_3$ photofragment kinetic energy distribution is unimodal with higher energy than available for the I* pathway, and accessing the $^1Q_1$ excited state is unlikely, being at higher energy\cite{jensen:2024}. Studies of CH$_3$I photodissociation at the longer wavelength region of the A-band region finds that the CH$_3$ photofragment internal energies differ for dissociation on the $^3Q_1$ pathway, exciting more in the $\nu_2$ `umbrella' mode as compared to the $^3Q_0$ and $^1Q_1$ dissociation pathways\cite{garcia-vela:2013}. Recent experimental results and calulations\cite{garcia-vela:2013,li:2005,rubio-lago:2011} have found that the $\nu_2$ CH$_3$ mode excitation distribution is quite constant for dissociation along the $^3Q_1$ dissociation pathway (i.e. $E_\textrm{{int}}$(CH$_3$) in Equ. \ref{Equ_2}), more strongly excited than for the $^3Q_0$ pathway accessed by photodissociation at the same energy.

The CH$_3$ photofragment kinetic energies in the EET dissociation features are remarkably similar for all of the substituted benzenes studied (Figs {\ref{fig_CH3I_on_mFBz_dFBz}} and {\ref{fig_CH3I_on_C6H5X}}). As seen in Fig. {\ref{fig_CH3I_on_mFBz_dFBz}}, the EET feature peak is consistently found to be slightly faster (by 1--2$\mu$s) for CH$_3$I on fluorobenzene thin films as compared to the difluorobenzene thin films. This could be the outcome of some steric effect for the CH$_3$I binding sites on the molecular films or could be due to a slightly higher electronic energy transfer from the fluorobenzene thin film. For fluorobenzene, the gas-phase fluorescence edge and continuum is at $\sim$7nm shorter wavelength\cite{fukuzumi:1991} (0.11eV higher energy as per Table {\ref{table_1_energies}}) than for difluorobenzene, of which the majority should be in the form of CH$_3$ photofragment kinetic energy (Equ. {\ref{Equ_2}}), which can account for the difference in peak positions observed. The variations in $S_1$ band edge energies for the other substituted benzenes do not fully align with the small differences in the CH$_3$ photofragment energies that are observed. For example, phenylacetylene has the lowest $S_1$ energy of those studied here, but the CH$_3$ photofragment kinetic energy peak in Fig. {\ref{fig_CH3I_on_C6H5X}}(c) is similar to the others in Fig. {\ref{fig_CH3I_on_C6H5X}}, so the systematic impact from this factor is not clear.

The energetics described in Section \ref{CH3X_energetics} can be used to estimate the initial excitation energies ($E_{exc}$ in Eq. {\ref{Equ_2}}) that produce the observed EET features. From the I and I* channel CH$_3$ kinetic energies (e.g. Fig. {\ref{fig_CH3I_on_pFBz}}), CH$_3$ photofragment internal energies are estimated to be $\sim$0.17eV and 0.09eV for the I and I* channels, albeit with rather large relative errors. These internal energies are similar to those seen in gas-phase CH$_3$I photodissociation at 248nm\cite{vanveen:1984,garcia-vela:2013} and consistent with finding somewhat more internal energy (mainly in the CH$_3$ $\nu_2$ excitation) in the I than in the I* channel. Using our observed CH$_3$ photofragment kinetic energies from neutral photodissociation as a reference (Fig. {\ref{fig_CH3I_on_pFBz}}), the EET peak for benzene suggests an excitation energy of 4.52eV, and an excitation energy of 4.35eV for CH$_3$I on the fluorobenzene thin film. For CH$_3$I on the other substituted benzenes, an initial excitation energy of approximately 4.24eV is estimated. For the CH$_3$I on naphthalene and fluoronaphthalene, an excitation energy of 3.88eV is suggested. These energies are below the respective $S_1$ 0-0 transition energies (Table {\ref{table_1_energies}}), lying within the range of $S_1$ fluorescence emission seen for these aromatic molecule systems. These energy estimates  are made assuming that the internal energies (non-translational degrees of freedom such as $\nu_2$ umbrella stretch mode of CH$_3$) are similar, though such energy changes would be small ($h\nu_2$=0.075eV\cite{li:2005}) unless the vibrational excitation distribution changes markedly. The apparent dominance of dissociation via the $^3Q_1$ state via EET for these excitation energies is somewhat different than the photoexcitation in the A-band described in Section \ref{CH3X_pdissn}. In photodissociation of gas-phase CH$_3$I, the $^3Q_0$ excitation is dominant in the low energy end of the A-band until the $^3Q_1$ become dominant below $h\nu \sim$3.90eV\cite{rubio-lago2009} and at slightly higher energy for CF$_3$I\cite{townsend:2004}. Likely this difference reflects the different transition probabilities for EET from the aromatics versus the photon transition probabilities. A similar assignment to $^3Q_1$ excitation has been made in CH$_3$I:O$_2$ cluster photochemistry\cite{bogomolov:2016}, where unimodal photofragment TOF spectra for the CH$_3$I dissociation were observed. 

A similar analysis of the photofragment energies from the CF$_3$I dissociation is more difficult due to the differing dynamics and larger uncertainty for the internal photofragment energies and surface physisorption well for the CF$_3$. The estimates for the initial excitation energies based on the TOF spectra such as in Fig. {\ref{fig_CF3I_on_various}} are systematically lower than the energy estimates from CH$_3$I on each corresponding aromatic thin film. Notwithstanding, it seems most likely that the initial excitation energies are similar for both CF$_3$I and CH$_3$I on each of the aromatic films. The origin of this difference could well be that the CF$_3$ photofragment kinetic energy distributions are shifted toward lower energies by interactions during the dissociation and desorption process, which might also underlie why the I and I* pathways are less well resolved in the adsorbed state TOF spectra (Fig. {\ref{fig_CF3I_on_various}}) than in gas-phase studies of CF$_3$I photodissociation.

\section{Summary and Conclusions}
Observation of the photodissociation dynamics for CH$_3$I and CF$_3$I adsorbed on thin films of aromatic molecules reveals that photoabsorption by the aromatic followed by EET can be a significant pathway for dissociation. Using 248nm incident light, benzene and many of the substituted benzenes display dynamics that reflect photoabsorption in the aromatic thin film characteristic of the $S_1$ excited singlet state, followed by EET and dissociation of the coadsorbate. For phenylacetylene and the naphthalenes studied, the initial photoabsorption occurs via the higher cross section $S_2$ excited state, which then undergoes rapid internal conversion to the longer-lived $S_1$ excited state, from which EET to the coadsorbates occurs. The photofragment spectra for CH$_3$/CH$_3$I and CF$_3$/CF$_3$I from the EET dissociation mechanism are unimodal, with energies determined from the $S_1$ excitation energies rather than the initial photon energy. The EET dissociation for CH$_3$I and CF$_3$I adsorbed on these aromatic thin films appears to proceed via the $^3Q_1$ excited state, a distinct difference from photodissociation of these species in this energy region.

\section*{References}
\bibliography{CH3I_Aromatics_EET_Bibliography}

\end{document}